The temperature-dependent chiral-induced spin selectivity effect:
Experiments and theory

Tapan Kumar Das,[1] Francesco Tassinari,[1] Ron Naaman,[1*] and Jonas Fransson[2*]

1) Department of Chemical and Biological Physics, Weizmann Institute, Rehovot 7610001, Israel
2) Department of Physics and Astronomy, Uppsala University, Uppsala 752 36, Sweden

**Abstract**

The theoretical explanation for the chiral-induced spin selectivity effect, in which electrons' passage through a chiral system depends on their spin and the handedness of the system, remains vague. Although most experimental work was performed at room temperature, most of the proposed theories did not include vibrations. Here, we present temperature-dependent experiments and a theoretical model that captures all observations and provides spin polarization values that are consistent with the experimental results. The model includes vibrational contribution to the spin orbit coupling. It shows the importance of dissipation and the relation between the effect and the optical activity.

**Introduction**

It is by now well established that when electrons move through a chiral electrostatic potential, one spin is preferred. The preferred spin depends on the handedness of the potential, so that electrons with spin aligned parallel to the velocity move more efficiently than electrons with spin aligned antiparallel to the velocity, for one handedness; the opposite will be true for the opposite handedness. This is termed the chiral-induced spin selectivity (CISS) effect [1].

The mechanism underlying the CISS effect was elusive. Although many of the theoretical treatments qualitatively provide the effect and indicate that indeed, the helical form of the electrostatic potential is responsible for spin-selective transport, quantitatively, however, most calculations failed to show quantitatively the magnitude of spin selectively observed experimentally. The failure of the models was associated with the small spin orbit coupling (SOC) typical of hydrocarbons and of the carbon atom itself [2, 3]. It was suggested that to obtain quantitative agreement between the theory and the experiments, one has either to replace the spin selectivity with angular momentum selectivity and to associate the spin observed with the SOC in the leads [4,5,6] or one has to find an enhancement mechanism that will somehow magnify the effect of the small SOC [7,8,9].

Although most of the CISS-related experiments were performed at room temperature [1] the model calculations were usually conducted for pure electronic systems that did not include any temperature effects. In the meantime, it also became apparent that the *pure* zero temperature single channel models are not consistent with the experimental observations and that a discrepancy exists between the symmetry arguments raised, based on these models, and the observations [10]. One realized that despite that the pure models indicate that two point contact experiments are unable to measure the spin-polarization in the linear regime, various experiments showed that it is indeed possible [11]. In addition, it was pointed out that the symmetry of the magnetoresistance curves, which one can observe in the CISS experiments, is not consistent with that observed in the case of giant magnetoresistance or chiral-magneto measurements [12].

Recently, several model calculations have resulted in large spin polarization when polaron [13] or vibrational effects [14,15] were considered. In the meantime, accumulating evidence has indicated that the CISS mechanism must include some additional features, since a correlation was found between the extent of spin polarization, when conducting through chiral molecules, and their optical activity [16,17,18].

Temperature dependence of the CISS effect may provide essential insight into the mechanism. This is indeed the focus of the present work that shows a clear enhancement of the effect with increasing temperature and a very good quantitative fit between a model that includes the vibrational effect and all the experimental result.

**Experimental results**

The magnetoresistance (MR) measurements were performed with crossbar geometry on a $SiO_2$ wafer with asymmetric and symmetric devices. The devices are described in the supporting information (SI).

Two types of symmetric devices were used. The first design was applied for studying the α-helix oligopeptides and the second for ds-DNA. In the two designs, both electrodes contained a ferromagnetic Ni layer.

All electrical measurements were carried out within the cryogenics system (Cryogenics, Ltd). A magnetic field of up to 0.9 T was applied perpendicular to the sample plane and the resistance of the device was measured using the standard four-probe method. A constant current of 0.01 mA was applied and the voltage across the junction was measured. The current versus voltage (I-V) studies were also carried out in the same device with the magnet up and magnet down with a field of 1 T perpendicular to the device. Two types of monolayers of chiral molecules were investigated. One is 40 bp ds-DNA with the sequence:

CGC TTC GCT TCG CTT CGC TTC GCT TCG CTT CGC TTC GCT T/3ThioMC3-D/
AAG CGA AGC GAA GCG AAG CGA AGC GAA GCG AAG CGA AGC G.

The other is oligopeptide, $SHCH_2CH_2CO$-$[Ala-AiB]_5$-COOH, when Ala is alanine and AiB refers to 2-aminoisobutyric acid.

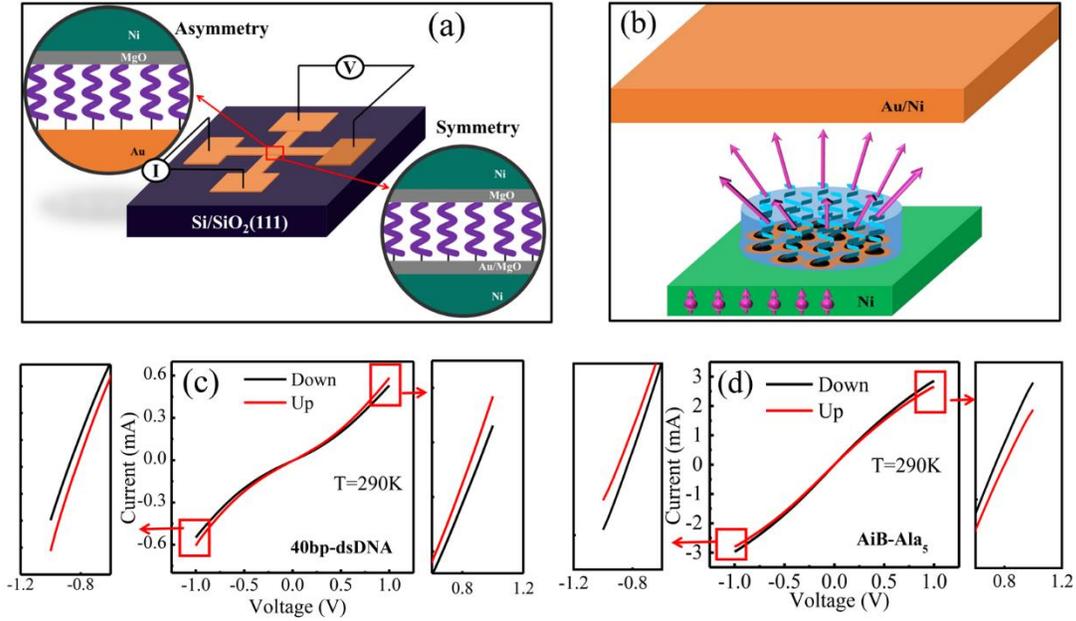

FIG. 1. The experimental set-up and current versus voltage measurements. (a) A schematic of a four-probe magnetoresistance (MR) device for both symmetric and asymmetric configurations with respect to the nickel electrode. Here, two types of symmetric device are shown. In type 1, the bottom electrode was made of nickel (40nm), followed by Au (5nm), and in the type 2, the bottom electrode was made of nickel (40nm), coated by MgO (2nm). Following the adsorption of the molecules on the bottom electrode, the top electrode was made of 2nm of MgO, followed by nickel and Au. (b) The typical trajectories of the electrons inside the device. Because of the large dimensions of the electrodes relative to the molecular size, electrons are collected from a wide angle, namely, also scattered electrons within the monolayer are collected. Hence, their spin polarization is reduced. (c,d) The current *vs* voltage (I-V) curves for symmetrical devices recorded at 290 K for (c) 40bp-long dsDNA and (d) [AiB-Ala]$_5$, when the current and magnetic fields are in same direction for magnet up (red) and in the opposite directions, for magnet down (black).

Fig. 1 (a) presents a scheme of the devices used. Fig. 1(b) shows the typical trajectories of the electrons inside the device. Note that because of the large dimension of the electrodes relative to the molecular size, the electrons are collected from a wide angle, namely, they scatter within the monolayer before reaching the electrode. Hence, their spin polarization is reduced. Consequently, the spin polarization measured in the micron size devices is consistently smaller than that measured using magnetic contact atomic force microscope (AFM) [19]. The current versus voltage curves for the case in which the electrodes are magnetized either UP or DOWN for symmetrical devices are shown in Fig. 1 (c,d). Very similar results were obtained for the asymmetric devices (see SI). One can observe that whereas for the DNA the current is higher when the magnet points UP, for the oligopeptides the current is higher when the magnet points DOWN. The temperature dependent current versus voltage is shown in Fig. S1. This finding is consistent with former studies and is correlated with the different handedness of the optical activity of the two molecules [19].

The temperature and voltage dependences of the spin-polarization (SP) for the systems studied are shown in Fig. 2. Fig. 2 (c,d) and Fig. S2 (a, c) shows the spin polarization as a function of temperature for the two molecules. Clearly the spin polarization increases with temperature and the increase does not significantly depend on the voltage applied between the two electrodes.

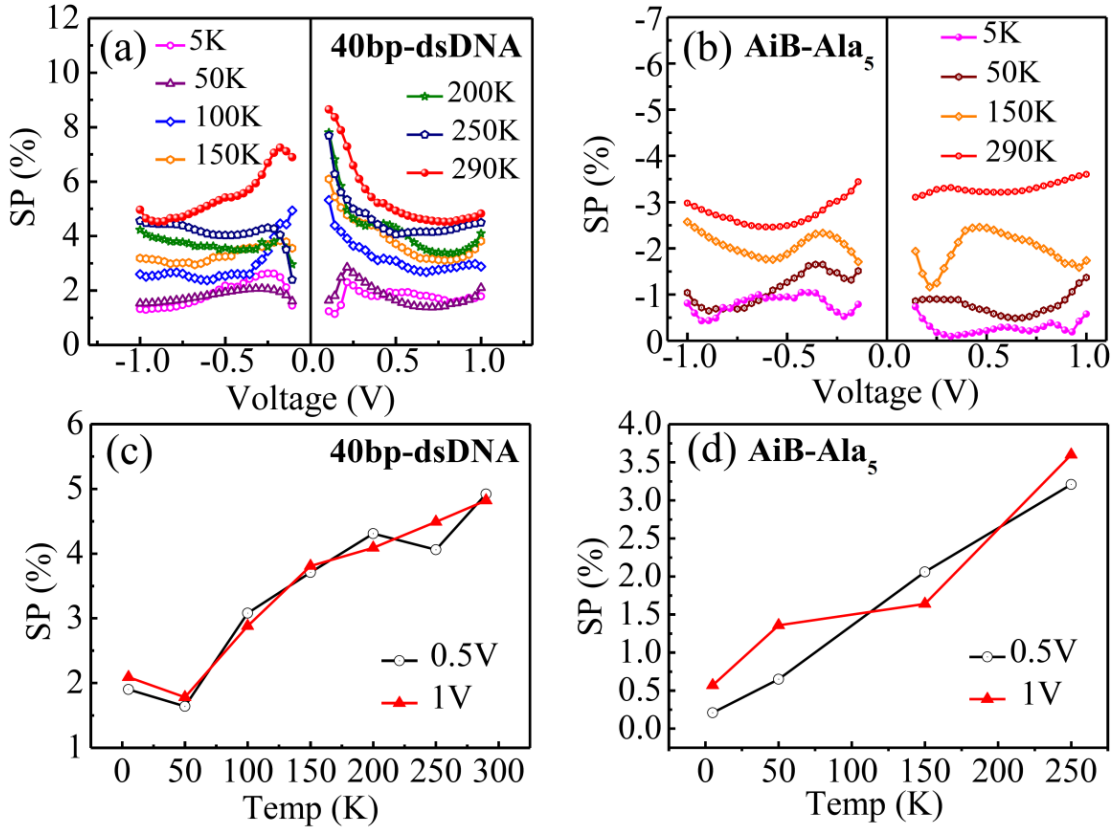

FIG. 2. Spin polarization (SP) calculated from the current-voltage (I-V) characteristics plot for magnet-up and magnet-down. (a) SP (in %) for the 40bp-length of dsDNA and (b) [AiB-Al]$_5$ oligopeptide measured at different temperatures. (c,d) The spin polarization as a function of temperature. SP = $(I_{up}-I_{down})/(I_{up}+I_{down}) \times 100$, where $I_{up}$ and $I_{down}$ are the current with spin aligned parallel and antiparallel to the magnetic field directions, respectively. The curves are given for the applied voltage 0.5V and 1V.

Figure 3 shows the magnetoresistance (MR) signal for symmetric devices. The MR curve is asymmetric and of an opposite shape, as indeed is expected with the CISS effect. The magnitude of the MR increases almost linearly with temperature. Similar results were obtained for asymmetry devices (Figs. S3, S4).

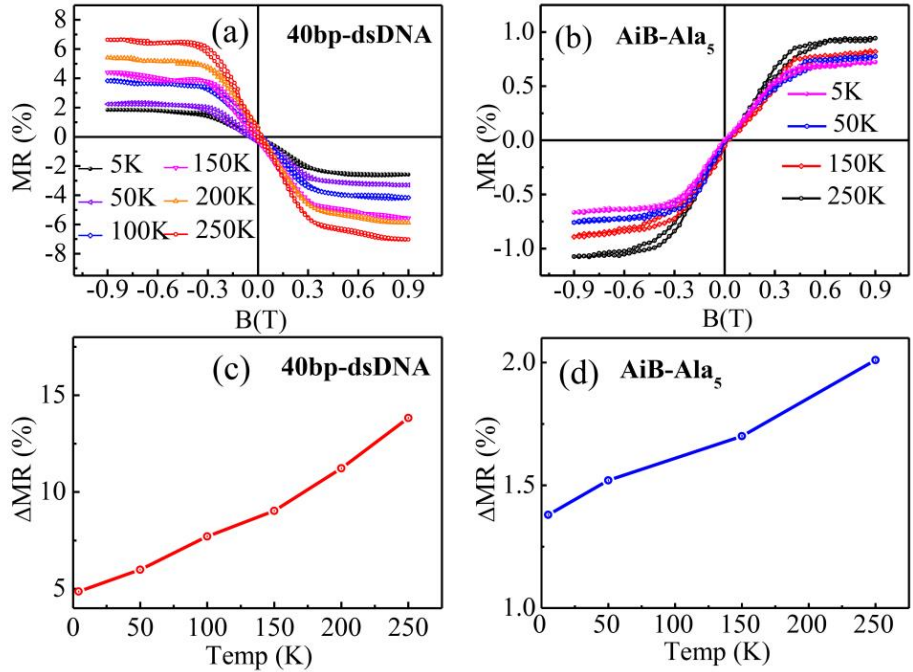

FIG. 3. Temperature-dependent magnetoresistance (MR) obtained with symmetrical devices. (a) A 40bp-length of dsDNA and (b) an [AiB-Ala]$_5$ oligopeptide measured at a different temperature with an input current of 0.01mA. (c) The ΔMR (%) values as a function of temperature for 40 bp—ds-DNA  and (d) for AiB-Al$_5$, where ΔMR (%)= MR(%)$_{0.9T}$+MR(%)$_{-0.9T}$.

The experimental results indicate that the spin polarization increases with increasing temperature. It also shows that in the MR studies, the high current is obtained when the magnet is pointing in the same direction, for the two current directions. This is contrary to the observations in the common Giant Magneto Resistance studies. These two phenomena cannot be explained with most theoretical models presented so far. In addition, in the CISS-related experiments, a correlation was found between the spin polarization and the optical activity [16,17,18], which also has remained unexplained. Next, we present a model that explains all the phenomena observed and provides an insight into why some of the former theoretical models have failed.

**The model:**

Here, we follow Refs. [15,20] and consider the role of vibrations in the CISS effect.

In order to illustrate the basic picture of the direction pursued here, we begin from the ordinary single electron Hamiltonian

$$H = \frac{p^2}{2m_e} + V(\mathbf{r}) + \xi(\nabla V(\mathbf{r}) \times \mathbf{p}) \cdot \frac{\hbar}{2}\boldsymbol{\sigma}, \qquad \text{Eq. (1)}$$

where $m_e$ is the electron mass, $V(\mathbf{r})$ is the effective confinement potential, and the last term defines the SOC contribution, $\mathbf{L} \cdot \mathbf{S}$, in terms of the orbital and spin degrees of freedom $\mathbf{L} = \xi \nabla V \times \mathbf{p}$ and $\mathbf{S} = \hbar \boldsymbol{\sigma}/2$, respectively, with $\xi = (2m_e c)^{-2}$. The operator $\mathbf{p}$ acts on everything to its right, whereas $\nabla$ acts only on the component directly adjacent to its right, such that $\nabla V \times \mathbf{p} = (\nabla V) \times \mathbf{p}$. Finally, $\boldsymbol{\sigma}$ denotes the vector of Pauli matrices.

When considering the role of the vibrations, the confinement potential $V(\mathbf{r})$ is developed by the vibrational coordinates $\mathbf{Q} = \mathbf{r} - \mathbf{r}_0$, where $\mathbf{r}_0$ denotes the equilibrium position, such that $V(\mathbf{r}) = V(\mathbf{r}_0) + \mathbf{Q} \cdot \nabla V(\mathbf{r})_{|\mathbf{r} \to \mathbf{r}_0} + \cdots,$. The orbital component of the SOC, $\mathbf{L} = \xi \nabla V \times \mathbf{p}$, can therefore, be written as $\mathbf{L}(\mathbf{r}) = \mathbf{L}_0 + \delta \mathbf{L}(\mathbf{Q}(\mathbf{r}), \mathbf{r})$, where

$$\mathbf{L}_0 = \xi \lim_{\mathbf{r} \to \mathbf{r}_0} \nabla V(\mathbf{r}) \times \mathbf{p}, \qquad \text{Eq. (2A)}$$

$$\delta \mathbf{L}(\mathbf{Q}(\mathbf{r}), \mathbf{r}) = \xi \lim_{\mathbf{r} \to \mathbf{r}_0} \nabla (\mathbf{Q} \cdot \nabla V(\mathbf{r}) + \cdots) \times \mathbf{p}. \qquad \text{Eq. (2B)}$$

Through this decomposition, we can define the static SOC $\mathbf{L}_0 \cdot \mathbf{S}$, which is the form that is normally considered in the absence of nuclear motion. Moreover, there is no immanent temperature dependence emerging from this contribution. By contrast, the correction to the static SOC, $\delta \mathbf{L} \cdot \mathbf{S}$, does provide a direct temperature dependence, arising from the coupling between the electrons and the nuclear vibrations. This can be understood by considering that the vibrational coordinate $\mathbf{Q}$ constitutes the nuclear displacement, which is thermally activated and becomes finite by anharmonicity (see for instance, Ref. 21).

The term $\delta \mathbf{L} \cdot \mathbf{S}$ represents the vibrationally assisted correction to the SOC, and defines the spin-dependent coupling between the electrons and nuclear motion, arising, as argued above, from developing the SOC in the vibrational coordinates. In the past, this contribution has traditionally been discarded; however, as is evident in the case in which the vibrations possess angular momentum, like in chiral molecules, there is no fundamental reason for such neglecting. For instance, under conditions where the equilibrium confinement potential, $V_0$, is close to a local extremum, the gradient $\lim_{r \to r_0} \nabla V(r)$ nearly vanishes, leading to a small SOC. Simultaneously, the second derivative, $\lim_{r \to r_0} \nabla [Q \cdot \nabla V(r)]$, may be substantial, which, in turn, leads to the vibrationally assisted correction to the SOC, being even the dominating contributor to the total SOC.

Here, we consider the temperature dependence as a result of coherent molecular vibrations, which is why we express the displacement operator $Q$ in the second quantization as $Q = \sum_\mu l_\mu \epsilon_\mu (a_\mu + a_\mu^\dagger)$, where $l_\mu = \sqrt{\hbar/2\rho v \omega_\mu}$ defines a length scale in terms of the density of vibrations $\rho$, a system volume $v$, a vibrational frequency $\omega_\mu$, and a polarization vector $\epsilon_\mu$, whereas $a_\mu + a_\mu^\dagger$ represents the displacement quantum at $r$ in the mode $\mu$.

In constructing our model, we assert that the displacement operator $Q$ defines a normal coordinate associated with the coupled equations of motion for the nuclei in the chiral molecule. Hence, the quantum operators $a_\mu$ and $a_\mu^\dagger$, respectively, annihilate and create a coherent molecular vibration in mode $\mu$. Consequently, the polarization vector $\epsilon_\mu$ carries the chiral symmetry of the molecule. Therefore, it enables a viable coupling between the angular momentum of the vibrations and the electron spin.

Next, we write Eq. (1) in the second quantization as

$$\mathcal{H} = \sum_m \psi_m^\dagger E_m \psi_m + \sum_\mu \omega_\mu a_\mu^\dagger a_\mu + \sum_{mm'\mu} \psi_m^\dagger U_{mm'\mu} \psi_{m'} (a_\mu + a_\mu^\dagger), \qquad \text{Eq. (3)}$$

where $E_m = \int \phi_m^* (p^2/2m_e + V(r_0) + L_0 \cdot S) \phi_m dr/v$ defines the single-electron energy matrix, including the static SOC, in terms of the eigenstates $\phi_m = \phi_m(r)$, and $\psi_m$ is the electron spinor, whereas the couplings between electrons and vibrations are defined through $U_{mm'\mu} = U_{mm'\mu} + J_{mm'\mu} \cdot \sigma$, where

$$U_{mm'\mu} = l_\mu \int \phi_m^* \epsilon_\mu \cdot \nabla V(r - r_0) \phi_{m'} \frac{dr}{v}, \qquad \text{Eq. (4a)}$$

$$J_{mm'\mu} = \frac{l_\mu \xi \hbar}{2} \int \phi_m^* \nabla \{[\epsilon_\mu \cdot \nabla V(r - r_0)] \times p\} \phi_{m'} \frac{dr}{v}. \qquad \text{Eq. (4b)}$$

Here, $U_{mm'\mu}$ defines the coupling between the electronic charge and the molecular vibrations, whereas $\bm{J}_{mm'\mu}$ denotes the strength of the vibrationally assisted correction to the SOC.

The roles played by $U_{mm'\mu}$ and $\bm{J}_{mm'\mu}$, in the physics of the structure, become clear by employing the canonical transformation $\tilde{\mathcal{H}} = e^S \mathcal{H} e^{-S}$, where the generating operator $S = -\sum_{mm'\mu} \psi_m^\dagger \bm{U}_{mm'\mu} \psi_{m'} (a_\mu - a_\mu^\dagger)/\omega_\mu$. Without loss of generality, we retain only a single vibrational mode, $\omega_0$, for simplicity. In the limit $\bm{J}\cdot\bm{\sigma} \to J\sigma^z$, the transformation enables the exact decoupling between Fermionic and the Bosonic degrees of freedom [15], such that $\tilde{\mathcal{H}} = \sum_m \tilde{\mathcal{H}}_m + \sum_{m\neq m'} \tilde{\mathcal{H}}_{mm'} + \omega_0 a_0^\dagger a_0$, where ($U_m = U_{mm0}$, $J_m = J_{mm0}$)

$$\tilde{\mathcal{H}}_m = (E_m - \frac{U_m^2 - J_m^2}{\omega_0}) n_m - 4\frac{U_m J_m}{\omega_0} s_m^z - 2\frac{U_m^2 - J_m^2}{\omega_0} n_{m\uparrow} n_{m\downarrow}, \qquad \text{Eq. (5a)}$$

$$\tilde{\mathcal{H}}_{mm'} = -\frac{2}{\omega_0}(U_m U_{m'} n_m n_{m'} - 4 U_m J_{m'} n_m s_{m'}^z - 4 J_m J_{m'} s_m^z s_{m'}^z), \qquad \text{Eq. (5b)}$$

given in terms of $n_{m\sigma} = \psi_m^\dagger(\sigma_0 + \sigma_{\sigma\sigma}^z \sigma^z)\psi_m/2$, $n_m = \sum_\sigma n_{m\sigma}$, and $s_m^z = \psi_m^\dagger \sigma^z \psi_m/2$.

In this form, it becomes clear that the coupling $U_m$ defines a vibrationally assisted electron-electron interaction, the third and first terms in Eqs. (5a) and (5b), respectively. The coupling $J_m$, on the other hand, defines an effective exchange interaction between different spins, the last term in Eq. (5b). Both contributions are viable mechanisms for enhancing the magnetic response in structures with broken spin symmetry. However, whereas a broken spin symmetry may be enhanced by a pure electron-electron interaction, it is largely defined by the asymmetry between the spin densities, whereas the energy difference between the spin states may be negligible. Such a density asymmetry is expected to provide a significant response in transport only at low temperatures, for example. In particular, the on-site interaction enhances the breaking of the spin symmetry in the chiral molecule and leads to a non-negligible spin selectivity, which was demonstrated in Ref. [9]. By contrast, the effective exchange interaction may induce a splitting between the energies of the electronic spin states. Hence, the effect of the redistribution of the spin densities is not only associated with different occupation probabilities for the spin states.

This mechanism is therefore expected to dominate the response at higher temperatures. The combination of the two vibrationally assisted couplings was previously shown to generate a strong CISS effect [15].

We now address the temperature dependence of the vibrationally generated interactions in the context of CISS, by employing a tight-binding model of the chiral molecule previously introduced in Refs. [15,20], represented by $\mathcal{H}_{mol} = \mathcal{H}_0 + \mathcal{H}_{vib}$, see the SI for details.. Here, $\mathcal{H}_0$ is the static contribution to the electronic energy, comprising the electron levels $\varepsilon_m$, the nearest-neighbor interaction, $t_0$, and static SOC through the next nearest-neighbor interaction, $\lambda_0$. Furthermore, $\mathcal{H}_{vib}$ comprises both a spin-independent vibrationally assisted nearest-neighbor interaction, $t_1$, and a vibrationally assisted next nearest-neighbor SOC, $\lambda_1$. In reference to the above discussion, the parameters $t_1$ and $\lambda_1$ represent the interactions $U_{mm'\mu}$ and $J_{mm'\mu}$, respectively. Upon mounting the model of the chiral molecule between metallic leads, of which one is ferromagnetic with a spin polarization parametrized by $p_L \in [-1,1]$, the charge is redistributed in the molecule, which is accompanied by the emergence of a finite spin-polarization [15,20]. This induced magnetic moment changes upon variation of the polarization $p_L$ which, thereby, changes the charge current through the structure. The procedure of how the charge current is calculated is thoroughly discussed in Refs. [9,15]. We calculate the charge current $I_\pm$ for a given spin polarization $p_L = \pm 0.5$ in the ferromagnetic lead.

In Fig. 4 we plot (a) the charge current for the bias voltage $V = 0.6$ V and (b) the spin selectivity for the vibrational frequency $\omega_0 = 4$ μeV (see the figure caption for other parameters). A similarity in the charge currents is the overall decay with increasing temperature. More striking, however, is that the decay strongly depends on the sign of the spin polarization, $p_L$. This effect results from the spin-dependent dissipation introduced by the presence of vibrations. Charge currents for bias voltages $V = 0.2$ V and $0.4$ V are shown in Fig. S5.

Molecular vibrations are a source of dissipation, which typically decrease the charge current as a result of inelastic scattering off vibrational excitations. At low temperatures, the number of such excitations is low, which leads to a small effect of the inelastic scattering. However, with increasing temperature, vibrational excitations occur; consequently, this enhances the effect of inelastic scattering. Therefore, the current decreases with increasing temperature.

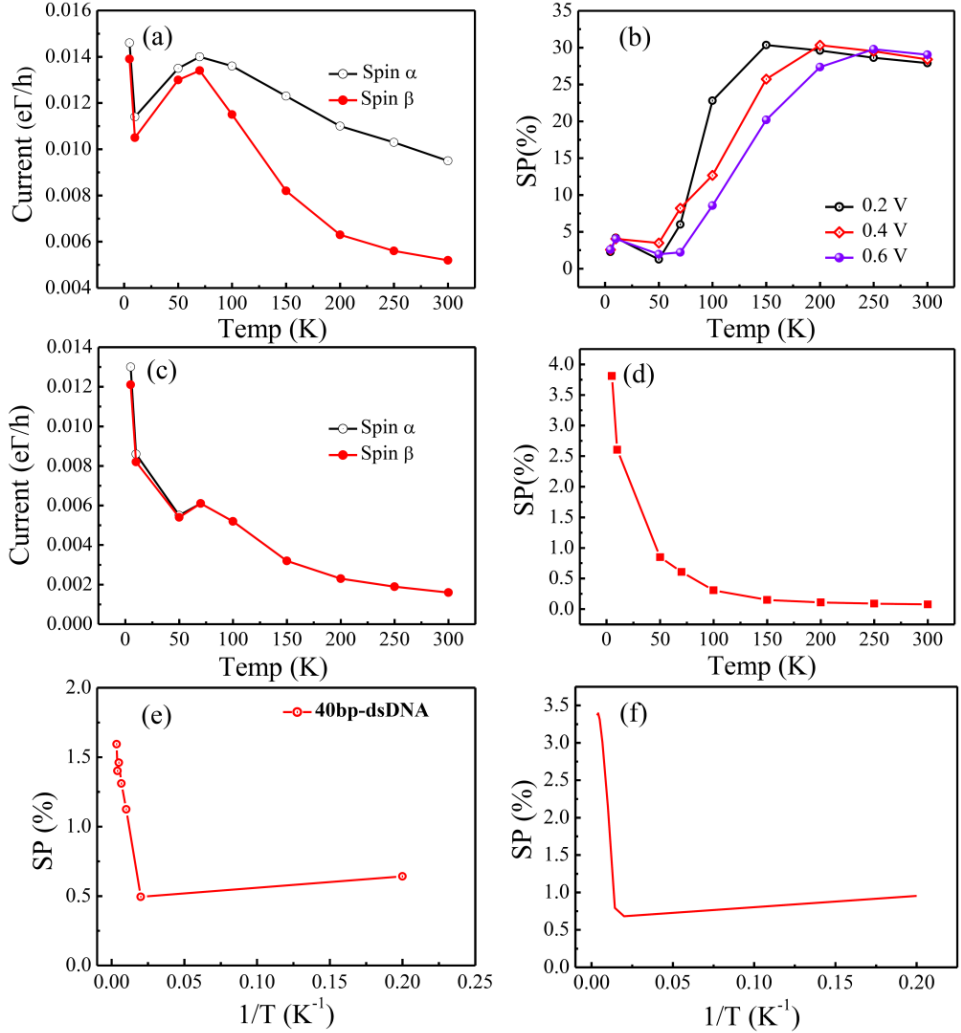

FIG. 4. (a,c) Simulated charge currents under polarizations, $p_L = \pm 0.5$ in ferromagnetic lead, and (b,d) the corresponding spin selectivity. The parameters used are $t_0 = 40$ meV, $\lambda_0 = t_0/40$, $t_1 = t_0/10$, $\omega_0 = t_0/10,00$, $\gamma = t_0/10$, and $V = 0.6$ V. However in (a,b) $\lambda_1 = t_0/400$) and in (c,d) $\lambda_1 = 0$ (c,d), Namely in c,d there is no contribution from the vibrations to the spin part of the vibrational enhanced SOC (Eq. 5b). In (e,f) there is comparison between the experimental results and the model calculations. (e) Ln of SP(%) versus 1/T is shown for 40b-dsDNA for an applied voltage of 0.5 V. (f) Model calculations Ln of SP(%) versus 1/T with parameters as in (a,b).

The source of the spin-dependent dissipation can be traced back to the vibrationally assisted SOC. We will refer to the processes captured in our model. To the second order in the interactions between sites, there are three different vibrationally assisted processes [15,20], namely (i) on-site and the second neighbor hopping, (ii) on-site and the fourth neighbor SOC, and (iii) the first and third neighbor mixed hopping and SOC. The first type of process is proportional to $t_1^2 \Sigma_m$, where $\Sigma_m$ is the electron-phonon interaction loop at the site $m$; it contributes to spin symmetrically to the dissipation. The second contribution is proportional to $\lambda_1^2 v_m^{(s)} \cdot \sigma v_{m+2s}^{(s')} \cdot \sigma \Sigma_m$, where the (unit) vector $v_m^{(s)}$ defines the chirality, and $s, s' = \pm 1$. Then, the processes that are on-site are spin symmetric, which can be understood by noting that the chirality vectors in this case, $v_m^{(\pm)}$ and $v_{m\pm 2}^{(\mp)}$, are parallel (see SI). The fourth neighbor SOC, nonetheless, breaks the spin symmetry, since the involved chirality vectors here are not parallel. However, the effect is the second order in the SOC parameter $\lambda_1$, which necessarily is small in perturbation theory. By contrast, the third types of processes, the first and third mixed hopping and SOC, which are proportional to $it_1 \lambda_1 v_m^{(s)} \cdot \sigma \Sigma_m$ and $it_1 \lambda_1 v_{m+s}^{(s')} \cdot \sigma \Sigma_m$, respectively, provide vibrationally assisted SOC to the first order in $\lambda_1$ and break the spin symmetry; spin splitting the electron energies. The temperature dependence of $\Sigma_m$ is dominated by thermally occupied vibrational excitations (Bose-Einstein distribution). Due to the scaling factors $t_1^2$ and $t_1 \lambda_1$, the impact on the electronic structure of the molecular vibrations is, however, shifted to higher energies than what is nominally suggested by the vibrational frequency. Hence, for $\omega_0 = 4$ μeV and the other parameters used here, a significant spin asymmetry between the currents would be expected to begin between 50 and 100 K, which is also seen in Fig. 4 (a). That the temperature dependence of the break in the spin symmetry is associated with mixed hopping and SOC processes is demonstrated in Fig. 4(c), where the two spin currents are calculated in the absence of the vibrationally assisted SOC. With increasing temperature, the two currents become increasingly equal, which is also confirmed by the decreasing SP (Fig. 4 (d)). The remaining spin asymmetry is attributed to the vibrationally induced electron-electron interaction, c.f., Eq. (5a).

The non-monotonic behavior in both the current and the spin selectivity is a signature of competing effects, such that the spin selectivity increases as long as the spin-dependent processes dominate dissipation, which is the case for temperatures below some threshold temperature, $T_c$, above which the spin symmetric processes become the dominating source for dissipation, such that the spin selectivity gradually decays.

As shown in Eqs. (5a), (5b), the vibrationally induced electron-electron and exchange interactions scale as $1/\omega_\mu$. Hence, we should expect that the vibrational effects on the break in the spin symmetry would be less effective, the higher the frequency of the vibrations. This is indeed observed in Fig. S6 where we repeated the simulations using $\omega_0 = 40$ μeV. Neither the spin-dependent dissipation nor the spin selectivity is as strong as in the case shown in Fig. 4, which can be directly linked to the one order of magnitude weaker exchange interaction energy.

In Fig. 4(e), the spin polarizations are presented on a logarithmic scale versus $1/T$, namely, an Arrhenius plot. Here, it is clear that a barrier exists for the increased spin polarization; it corresponds to a temperature of about 50K about 5 meV for DNA. These barriers correspond to vibrational frequencies of about 30-40 cm$^{-1}$ (4 meV), namely, about 1 THz. Hence, they must refer to vibrations that include a large part of the whole oligomer.

Using the parameters of the model, we reproduced the experimental results with excellent agreement, as shown in Fig. 4 (f). It clearly shows large spin polarization at high temperatures (low $1/T$ values); then there is a drop, but upon further cooling, the spin polarization increases slightly. This increase results from inelastic scattering due to the electron-vibration interaction, which is described in Eq. (5a). Upon cooling, this inelastic process decreases and the spin polarization that results from the static spin orbit coupling appears. We wish to emphasize that the model did not attempt to present precisely the experimental results, and it does not correctly represent all the molecular features. However, apparently it can describe the experimental results.

It is important to understand that the model relates the spin polarization to the molecular polarizability, through the polarizability term $\epsilon_\mu$ in Eq. (4). It is well established that the polarizability is proportional to the optical activity [22,23]. Hence the present model relates the spin polarization to the optical activity of the molecules. The model also indicates the importance of coupling with low-frequency vibrations/phonons, which means that the electron transmission through the molecules is not ballistic, but instead involves dissipation of energy and momentum. The model also result in the same symmetry in the I-V curves, as observed in all CISS results and which are opposite to the symmetry observed in common magnetoresistance devices (see SI and Fig. S7).

**Conclusions**

The experimental work presented here shows that the spin polarization, observed in α-helices, oligopeptides, and double-stranded DNA, results from the CISS effect, which increases with energy. The activation energy associated with the enhancement of the CISS effect corresponds to populating acoustic phonons in the systems. Hence, clearly the CISS effect is significant at room temperature. The theoretical model is based on enhancement of the SOC by the vibrations and the results obtained from the simple model coincide amazingly well with the experimental data. This is due to the contribution of the low-frequency modes, which are cooperative modes and therefore do not appreciably depend on the detailed molecular structure. The observations here are consistent with recent results obtained for inorganic crystals, where acoustic chiral phonons were shown to interact with the electrons′ spin and consequently, enhance the range of spin transport [24,25,26,27].

The experimental results, the model, and the fitting between them also explain why the zero temperature models fail to obtain the magnitude of the CISS effect and its qualitative behavior in terms of symmetry constraints. We believe that the present work now establishes the foundation to construct a more detailed model that will properly describe the molecular system and the vibrational effect.

**Acknowledgments**

RN acknowledges support from the MINERVA foundation and from the Israel Ministry of Science. J.F. acknowledges support from Vetenskapsrådet and Stiftelsen Olle Engkvist Byggmästare.